# Comparative Study of Magnetic Behaviour in Three Classic Molecular Magnets

Tanmoy Chakraborty,[1] Tamal K. Sen,[1] Harkirat Singh,[1] Diptaranjan Das,[1] Swadhin K. Mandal,[1] Chiranjib Mitra.[1*]

[1]*Indian Institute of Science Education and Research (IISER) Kolkata, Mohanpur Campus, PO: BCKV CampusMain Office, Mohanpur - 741252, Nadia, West Bengal, India.*

*chiranjib@iiserkol.ac.in

## ABSTRACT

We have studied the magnetic properties of three phenalenyl based organic neutral radicals. The first one is a Heisenberg chain antiferromagnet with one unpaired spin per molecule; second one is a diamagnetic, exhibiting a diamagnetic to paramagnetic phase transition at high temperature; the third one comprises of free neutral radicals and shows paramagnetic behaviour. Temperature dependent magnetic susceptibility measurements and isothermal magnetization measurements (as a function of magnetic field) were performed on all the three systems. In the case of the antiferromagnetic system, temperature dependent susceptibility and magnetization isotherms were fitted to the Bonner Fisher model. In the case of second system the diamagnetic to paramagnetic phase transition is investigated by performing isothermal magnetization measurements in the two different phases. The diamagnetic to paramagnetic phase transition seems to be of first order in nature.

# 1. INTRODUCTION

The study of magnetism is mostly concentrated on the materials which are inorganic in nature and atom-based. But some exciting and interesting properties lie in the study of physics and chemistry of molecular magnets which are chemically a lot more complicated than inorganic magnets [1]. These relatively new class of magnets provide a wide range of applications in many areas of science and technology. Molecular magnets are primarily organic in nature. Magnetism in a molecular magnet deals with isolated molecules or assemblies of molecules with one or more magnetic centre in a single molecule. As a result, the magnetic building block for a molecular magnet is molecules, instead of an atom. Intramolecular forces in these systems dominate over the intermolecular forces. Because of the fact that the intermolecular forces are non-covalent (hydrogen bondings, Van der Walls interactions, donor-acceptor charge transfer etc.) in nature. Consequently the crystals are relatively softer than the inorganic ones where ionic cores dominate. These weak intermolecular forces lead to sometimes special optical and magnetic properties. The physical properties of molecular magnets are determined by its crystallographic and electronic structure [2]. One important reason of studying molecular magnet is its extraordinary tunability. The flexibility of Tailor made chemical structures allows one to modulate and fine tune its solid state packing interaction structure and hence its physical properties. This way one can design one's own model system with desired magnetic properties. The fascinating features of these of systems have drawn the attention of many researchers.

In this work we have chosen three molecular magnets which are composed of phenalenyl-based neutral radicals [3, 4 and 5]. Three benzene rings fuse in a triangular fashion to form one phenalenyl basis. Presently we have focussed on different electronic spin structures and diversity in magnetic properties of these three radicals with same phenalenyl basis. Due to interesting spin structure, tunability and higher spin-relaxation time, these delocalised ply- based neutral radicals have tremendous impact in context of quantum information processing and molecular spintronics [6]. These molecular magnetic system could have very promising applications as entangling channel, which can be used as quantum networks connecting to quantum gates. These neutral organic molecules form molecular crystals with single molecular species that function like a mono-atomic metal [7, 8]. This means that radicals [A radical is a molecule with unpaired electron] arranged in arrays in a molecular crystal is equivalent to the atoms in a metal. That's how a lattice is formed where spins are arranged in a periodic manner these spitrobiphenyl boron radicals. These unpaired electrons serve as the charge carriers and orbital overlap between the radicals give rise to quarter-filled energy bands, which results in the systems to behave like a Mott insulator.

Out of these three materials the first system we considered is spiro-bis (1, 9-disubstituted-phenalenyl) boron neutral radical (see figure 1(a)), which henceforth will be referred to as the first system. Its molecular structure is shown in Figure 1(a) and is based on the diaminophenalenyl system [3]. It is known from the magnetic susceptibility measurements that these radicals exist as isolated free radicals with one spin per molecule. Earlier study has shown [3] that for this particular system the unpaired spins interact with other neighbouring spins along one particular direction by isotropic Heisenberg interaction. These spins are arranged in a chain like structure where intrachain overlap dominates over the interchain interactions. Therefore these systems behave like a spin chain and can be modelled by a Heisenberg Hamiltonian with nearest neighbour coupling. The interaction between one spin and its next nearest neighbour is so weak that it can be neglected. It so happens that antiferromagnetic interaction is favoured in this case. The general form of spin chain Hamiltonian with nearest neighbour Heisenberg interaction can be written as

$$H = 2J \sum_i \left[ \alpha S_i^z S_{i+1}^z + \beta (S_i^x S_{i+1}^x + S_i^y S_{i+1}^y) \right] \qquad (1)$$

Where J is the exchange integral and $S^x$, $S^y$, $S^z$ are the components of the total spin S along x, y and z direction respectively. When α = β = 1 the Hamiltonian takes the form of isotropic Heisenberg model. We have studied temperature (T) dependent magnetic susceptibility (χ) and have fitted the χ vs. T data with Bonner and Fisher model [9](see Fig. 1(b)). The value of exchange integral obtained is -16.6 K which is close to the value reported in the previous literature [3]. We have also obtained magnetization isotherms at various temperatures where we studied the variation of magnetization as a function of magnetic field. Bonner and Fisher carried out calculation on isotropically interacting spin chain systems containing from 3 to 11 spins and extrapolated it for infinite length chain with a good agreement [9]. We have generated numerical data for magnetization with varying magnetic field for 10 spin chain using Matlab. The theoretical curves are in very good agreement with the experimental data within experimental error.

The second system, that we have studied is butyl substituted N,O-donor based spiro-biphenalenyl radical (2, Figure 1(a)). In this system one can clearly see a phase transition from a diamagnetic to a paramagnetic phase in magnetic susceptibility vs. temperature data [4]. This transition temperature is also associated with the change in conductivity of the system. We have shown magnetization isotherms in these two different phases to bring out the distinct nature of the paramagnetic and diamagnetic phases. This system has already shown magneto-optical-electronic bistability with variation of temperature [10]. This interesting property of this material promises a potential application in molecular spintronics devices. The structure of the system is shown in Figure 1(a).

The third system we have attempted is N,O-donor based hexyl substituted spiro-biphenalenyl radical (3,figure 1(a)). This system consists of monomeric radicals with each molecule carrying a spin. Therefore they behave like free spins to exhibit paramagnetic Curie behaviour [5]. We have measured the magnetic susceptibility as a function of temperature. Our magnetic susceptibility data also captured the paramagnetic behaviour with some difference [11] from the existing literature [5] [†] and is reported in next section.

## 2. MATERIALS AND METHODS

We have synthesised and crystallised the first, second and the third samples in a single crystalline form as mentioned in the reference [3], [4], and [5] respectively. The generation of radical and crystallization were carried out inside a $N_2$ filled Glovebox using dry solvents. Subsequently we have measured magnetic properties of each of the three samples in a Quantum Design SQUID magnetometer.

For the first sample we have studied magnetic susceptibility in a temperature range of 2 K to 300 K. We have studied magnetization isotherms with varying magnetic field (H) from zero to 7 Tesla. This measurement was carried out at different temperatures from 2K to 30K. For the second sample, χ vs. T data were taken from 2 K to 360 K as we wanted to capture the diamagnetic to paramagnetic transition around 322 K and M vs. H data was taken at 300 K and 335 K. For the third sample only susceptibility measurement was done from 2 K to 40 K.

## 3. RESULTS AND DISCUSSION

We have performed zero field Magnetic susceptibility measurement on the first system in a temperature range of 2 K to 300 K (see Fig. 1(a)). Earlier it has been shown that for high temperatures greater than 35 K, the magnetic susceptibility data can be described by Curie-Weiss behaviour $\chi = C/(T-\Theta)$ with Weiss constant $\Theta = -14$ K and Curie constant $C =$ and 0.34 K*emu/mol [3]. The Curie constants are close to 0.375 K*emu/mol, the value expected for a neutral radical with one unpaired spin per molecule. This is also supported by the fact that the molecule does not pack within van der Waals radii of carbon atoms indicating non-interacting spins at least at higher temperature. However, the low-temperature data show significant deviations from Curie-Weiss behaviour due to antiferromagnetic coupling of the unpaired spins, with a maxima in the $\chi(T)$ curves at $T_{MAX} = 23$ K. For spin systems with isotropic Heisenberg coupling, Bonner and Fisher have derived zero field antiferromagnetic susceptibility for 3 to 11 spins and extrapolated for an infinite number of spins with a good agreement [9]. Hall [12] efficiently fitted the numerically evaluated data obtained by Bonner-Fisher to the following equation

$$\chi \approx \frac{Ng^2\mu_B^2}{k_BT} \cdot \frac{0.25 + 0.14995X + 0.30094X^2}{1.0 + 1.9862X + 0.68854X^2 + 6.0626X^3} \qquad (2)$$

Where $X = J/K_BT$ and the other symbols have their usual meaning. We have used this equation to get a reasonably good fit with our zero field magnetic susceptibility data. In this fitting we have used the intrachain exchange coupling J as a fitting parameter. We obtained J value of 16.6 K in units of $K_B$. The fit is shown in Figure 1(a) where the the solid line depicts the fitted curve and the open circles representing the experimental data.

We have investigated the behaviour of magnetization isotherms (of the first system) as a function of magnetic field for various temperatures. For our physical system there is one unpaired spin per molecular site which interacts with other neighbouring spins along one particular direction. As mentioned earlier this spin system behaves as a 1D spin chains. Therefore physically it has infinite number of interacting spins in one chain. For fitting the experimental magnetization curve one needs to derive an expression of magnetization for infinite number of spins. Since practically it's not possible to deal with infinite dimensional Hilbert space, Bonner and Fisher have numerically evaluated field dependent magnetization for N=10 spins and assured the convergence for N tending to infinity [9]. Based on this we have written a code in MATLAB for 10 spins to calculate magnetization isotherms as a function of field and fitted to our experimental data. Here we have considered the data set for T = 17.5 K which is below the antiferromagnetic ordering temperature. This fitting is shown in Fig 1(c) and it appears to be a good fit. The open circles represent experimental data and the fitted line is shown by the solid red line. The plot shown in Fig. 2(a) is a 3D plot which captures the variation of magnetization as a function of field and temperature.

In case of the second system we performed temperature dependent magnetic susceptibility measurement in the temperature range of 2 K to 360 K and is shown in Fig. 2(b). At low temperature regime (< 100 K) this system exhibits Curie behaviour while at a higher temperature (approximately 320K-340 K) the system undergoes a phase transition from diamagnetic to paramagnetic phase. Though the system exhibits paramagnetic behaviour in the temperature range of 2 K – 300 K this is in principle a diamagnet. The paramagnetic behaviour arises owing to the presence of some impurities [4]. Above 200 K it shows temperature independent diamagnetic behaviour up to 340 K. At 340 K the susceptibility shows a random jump and then exhibits a paramagnetic behaviour above this temperature. Upon cooling it continues in the paramagnetic phase down to 320 K, whereupon the susceptibility suddenly drops and continues in the diamagnetic phase as we further cool the system. The heating and cooling runs are indicated by arrows in the inset of Fig 2(b). One can clearly see a hysteresis in susceptibility as a function of temperature. This hysteresis accompanied by the sudden jump at 340 K in the heating curve and a sudden drop in susceptibility in the cooling curve at 320 K is indicative of a first order phase transition. With increasing temperature the coupled spins decouple to free spins. That is the reason the system undergoes a phase transition from diamagnetic $\pi$-dimer to paramagnetic $\pi$-dimer state [4]. Electrical and optical measurements on the same system have also shown this bistability when the temperature is varied [10].To demonstrate that the phase below 320 K is indeed a diamagnetic phase, we have carried out magnetization measurements at 300 K, which is just before the transition occurs. We have also taken magnetization isotherms at 335K, which is above the transition temperature. The former clearly shows a diamagnetic behaviour, whereas the latter magnetization isotherm depicts a paramagnetic behaviour. In the isotherm taken at 300 K, the small increase in magnetization at low field value is due to the paramagnetic impurities mentioned above. To fit the magnetization isotherm taken at 335 K fits, we excluded some low field data points to avoid the contribution from the impurities. We tried to fit the Brillouin function (equation no. 3) to the data (Fig. 2(a)), but it was not a satisfactory fit.

$$B_J(y) = \frac{2J+1}{2J}\coth(\frac{2J+1}{2J}y) - \frac{1}{2J}\coth\frac{y}{2J} \quad (3)$$

Where $y = g\mu_B JB/K_B T$ and the other symbols have their usual meaning. However, when this data was fitted to a straight line, we obtained a very good fit, which suggests that the system is a Pauli paramagnet [13]. The occurrence of the Pauli paramagnetic state in this temperature regime (above 320 K) suggests that there are free electrons in the system, whereas in the regime below 320 K, the system behaves like a diamagnet, suggestive of an insulating phase. The contribution to diamagnetism comes from ring currents arising from closed shells. This is a common feature in insulating systems. The presence of free electrons above 320 K indicates metallic behaviour, whereas below this temperature it exhibits insulating behaviour. Thus the system undergoes a metal-insulator transition around 320K. This behaviour has been captured earlier through conductivity measurements [10], which results from a structural phase transition around 320 K. We intend to explore this in more details in future.

Magnetic susceptibility vs. temperature data for the third system is shown in Fig. 3(b) in the temperature range 2 K to 40 K. This curve shows the signature of existence of paramagnetic state in this temperature range. The experimental data is fitted to the Curie function C/(T-Θ). The fit is shown in Fig. 3(b) and appears to be a good fit. The data is shown by the open circles and the fitted curve by the blue solid line. From this fitting the value of Curie-Weiss constant (Θ) is obtained 6 K. Therefore the radicals in this system exist as free radicals with one spin per molecule to show free spin paramagnetism.

## 4. CONCLUSION

This study was a part of understanding these molecular magnetic systems from the perspective of organic spintronics where one can have multiple applications of these systems either as an entangling channel for quantum computation or as molecular switches which can also be used as quantum gates. The first system in this sample exhibited antiferromagnetic ordering at 23K, which can have significant application in quantum networks, since any isotropic antiferromagnetic spin half system will exhibit entanglement well below the ordering temperature. And this system being one dimensional in nature can be used as quantum wires, which will teleport quantum information between two quantum gates. The second system which exhibits a temperature dependent transition between insulating diamagnetic state to a metallic paramagnetic state can be used as a molecular switch, where the system can be locally heated by an infrared laser thereby inducing the system from one state to another. This can have potential application in terms of fabricating both classical and quantum gates. The magnetization isotherms in the two regimes - paramagnetic metallic and diamagnetic insulating – show strikingly linear feature, showing that the paramagnetic regime is a rare Pauli paramagnetic one, quite unprecedented in organic systems. If one can switch the system's magnetic state under the application of a high magnetic field (say 5 Tesla), then there would be an enormous jump in the magnetization as the two regimes show linear dependence as a function of field but with slopes of different signs. This can help in fabrication in magnetic switches. The third system didn't show any striking feature and was a paramagnet in the entire temperature range. This result disagrees with previously reported data and we suspect that the antiferromagnetic phase reported in the previous case could result owing to the ordering of some impurities present in the system.

## 5. ACKNOWLEDGMENT

The authors would like to thank the Ministry of Human Resource Development, Government of India, for funding.

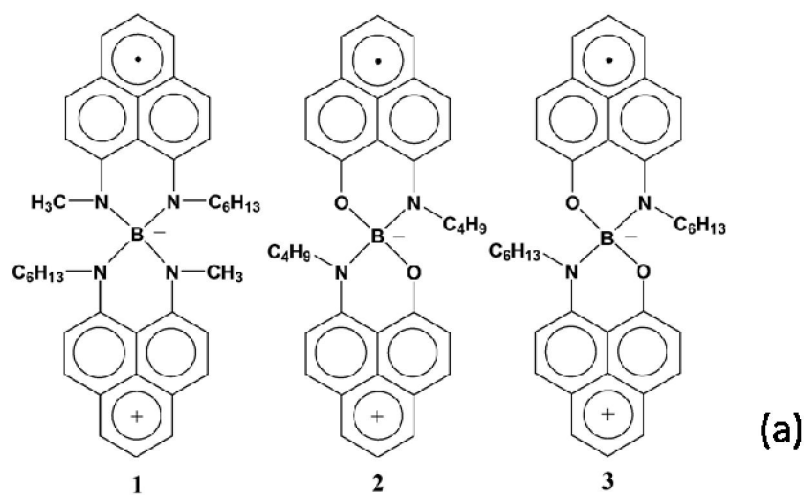

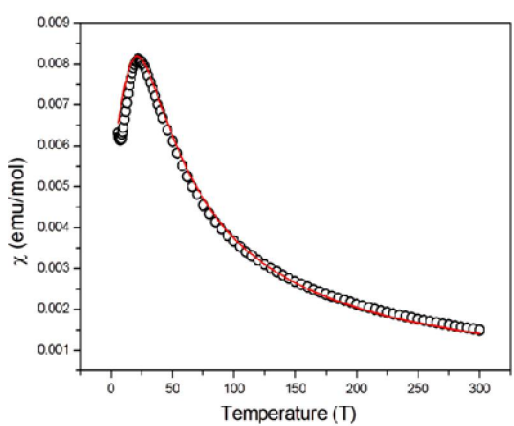     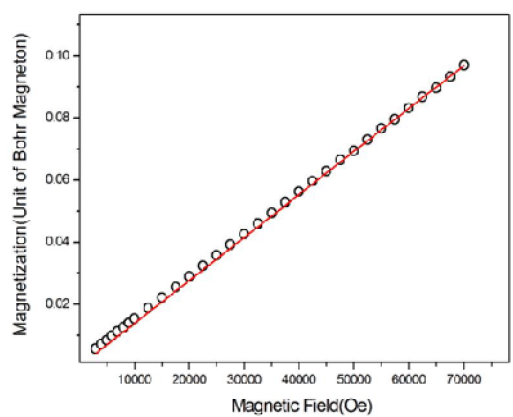

**Fig. 1.** (a) Molecular drawings of (1) first (2) second and (3) third systems drawn by the software CHEMDRAW. (b) Temperature dependent magnetic susceptibility of the first system. Open circles represent the experimental data and the fitted curve (Bonner-Fisher model) is shown by the solid red line. (c) Experimental data of magnetization collected at T = 17.5 K for the first system (open circles) as a function of magnetic field. It also shows the fit to the theoretical curve (solid red line) derived using the Heisenberg linear chain model.

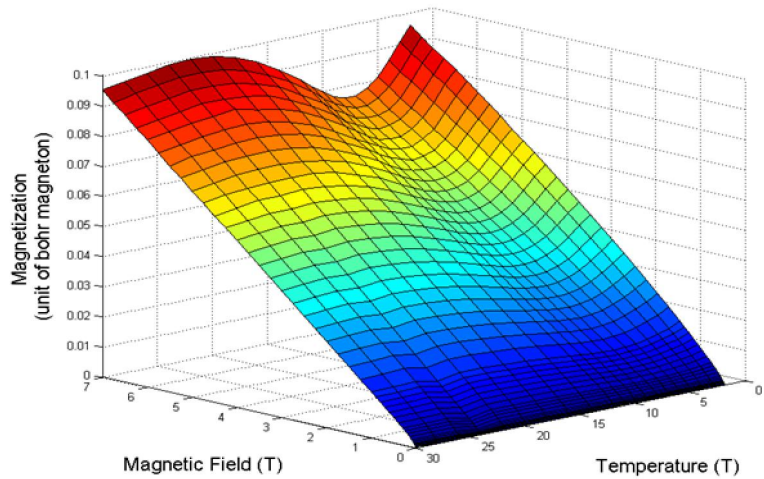

(a)

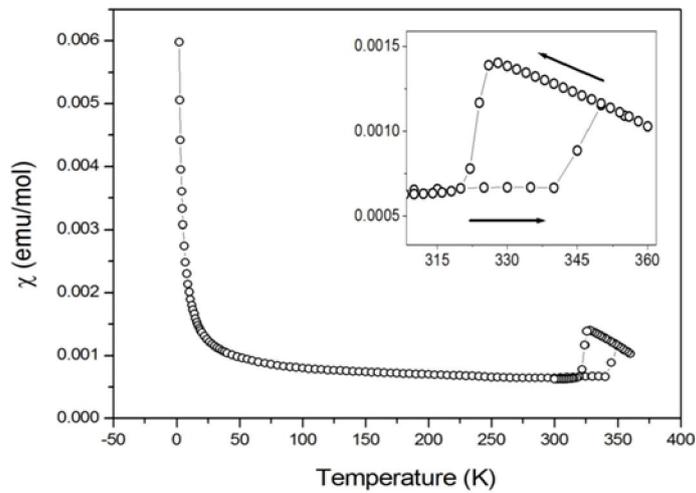

(b)

**Fig. 2.** (a) 3D plot showing Magnetization, magnetic field and temperature along the three axes for the first system. (b) Magnetic susceptibility as a function of temperature for the second system. Inset shows the expanded region to give emphasis on the hysteresis (see text).

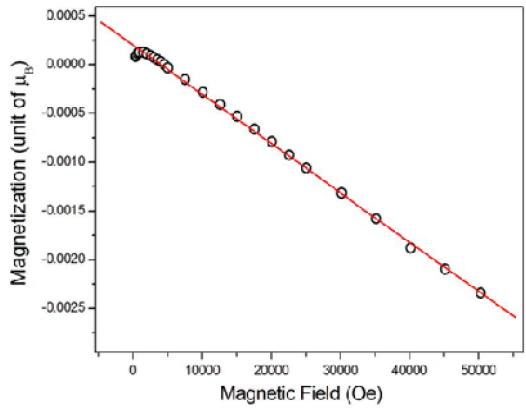 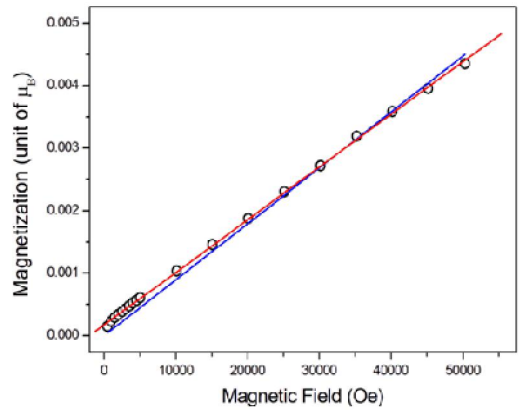

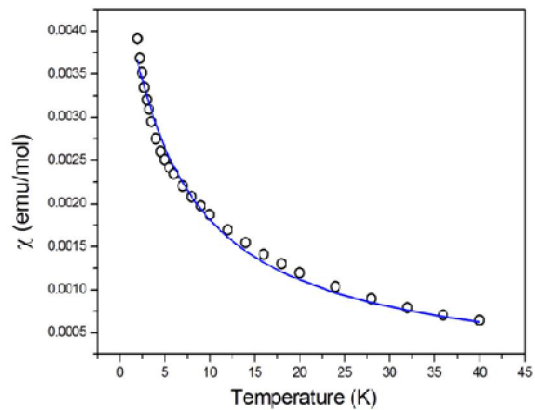

**Fig. 3.** (a) Field dependent Magnetization of the second system for a fixed temperature T = 300 K. The data points are shown by circles and the fitted curve (see text) is shown by the solid blue line. (b) Field dependent Magnetization of the second system for a fixed temperature T = 335 K. The data points are shown by circles. The red line is fit to the experimental data using Brillouin function and the blue line is linear fit to the data which represents Pauli paramagnetic behaviour. (c) Magnetic susceptibility of the third system. Open circles represent the experimental data and the solid blue line is the Curie-Weiss fit to the experimental data.